\documentclass[amsmath,amssymb,aps,prb,twocolumn,floatfix]{revtex4-1}
\usepackage{amsmath}
\usepackage{mhchem}
\usepackage{multirow}
\usepackage{xcolor}
\usepackage{braket}
\usepackage{hyphenat}
\usepackage{braket}
\usepackage{graphicx} 
\usepackage{dcolumn}
\usepackage{bm}
\usepackage{ragged2e}
\usepackage[caption=false]{subfig}
\usepackage{tabularx}
\usepackage{hyperref}
\usepackage{epstopdf}
\usepackage{epsfig}
\usepackage[utf8]{inputenc} 
\graphicspath{{./figures/}}

\begin{document}
\preprint{APS/123-QED}

\title{Relevance of Shockley states on the electrical and thermoelectric response of 
gold-based single-molecule junctions}

\author{Saúl Sánchez-González}
\author{Amador García-Fuente}%
\author{Jaime Ferrer }
\email{ferrer@uniovi.es}

\affiliation{
Departamento de Física, Universidad de Oviedo, 33007 Oviedo, Spain. \\ Centro de Investigación en Nanomateriales y Nanotecnología, Universidad de Oviedo–CSIC, 33940 El Entrego, Spain
} 

\date{\today}

\begin{abstract}
\noindent Noble metals break preferably exposing (111)-oriented surfaces, that host Shockley type surface
states (SSs). Nevertheless, the relevance of SSs on the electrical properties of gold-based molecular
junctions has not been explored in detail yet. Here, we present ab initio simulations that show how the
gold (111) SS, that lies approximately 0.5 eV below the Fermi energy, is key to determining correctly
the electrical and thermoelectric response of the above junctions. We show how the ability to shift in
a controlled way the energy position of gold SS enables us to tune the electrical and thermoelectric
response of gold molecular junctions. We also show that gold’s SS appears in our simulations only
if the 5d orbitals are included explicitly in the valence shell. To illustrate this behaviour, we discuss
in detail Benzenediamine (BDA) and Benzenedicarbonitrile (BDCN) gold (111) junctions.
\end{abstract}

\maketitle

\section{Introduction}
Energy supply and sustainability is one of the biggest
problems of today’s society. The decreasing amount of
fossil fuels and the growing energy demand have caused a
huge raise in electricity prices. Apart from that, the large
amount of CO$_2$ emitted in fossil fuel based energy production is one of the main responsibles of worldwide climate change. In order to address these challenges, renewable energy sources should be developed and energy demand reduced. One of the best approaches to these problems are thermoelectric devices which can convert waste heat (e.g. car exhaust pipes, house chimneys, industrial synthesis) directly into electricity.\cite{Warming1,Warming2} The efficiency of a thermoelectric 
material is determined by the figure of merit $ZT=GS^2T/\kappa$, a dimensionless magnitude where $G$ is the 
electronic conductance, $T$ is the temperature, $\kappa$ is the thermal conductance (with contributions 
from both electrons and phonons) and $S$ is the Seebeck coefficient (also known as thermopower). \cite{ZT1,ZT2}

A large value of $ZT$ is needed in order to achieve 
efficient energy conversion (ideally $ZT\geq$ 2 to compete 
with standard energy conversion methods). \cite{Efficiency1,Efficiency2}
In the search for systems with the adequate $ZT$ values, several strategies such as 
alloying, doping and band engineering\cite{MeritStrategies,doping1,doping2} 
have been used to increase it without much success, so a completely new approach may be needed.

Recently, experimental studies have shown that 3D topological insulators (TI) such as $\ce{Sn-Bi_{1.1} Sb_{0.9}Te S_{2}}$
(Sn-BSTS) may achieve a large enhancement in $ZT$\cite{SurfaceStateEnhancement} 
due to the presence of surface states.
Even though it is still under discussion whether all surface states could be utilized to improve 
the thermoelectric performance, simpler and previously known topological insulators are now being 
revisited.\cite{revisit} One of the most striking results of this research is that the well-known Shockley 
surface states (SSs) at noble metal (111)-oriented surfaces have the same properties as the ones seen in TIs.\cite{GoldTopological} Despite the fact that SSs have been studied in depth in 
surface physics,\cite{SS1,SS2} their impact in the electrical and thermoelectric response of gold-based 
molecular junctions has been largely overlooked. 

We show here that SSs are key to determining correctly the conductance and thermopower 
response of gold-based single-molecule devices since they feature a narrow peak right 
below the Fermi energy\citep{Papior} not only in the surface density of states, but also in the electrical
transmission function. SSs survive at gold's (111) surface after the adsorption of different
adsorbates like alkali metals,\citep{alkali1,alkali2} guest noble metals\citep{noblemetal1}
or CO,\citep{CO1} merely shifting their energy. In other words, adsorbates play the role
of dopants for gold (111) SSs.

The key role of SSs in the transmission of gold-based molecular junction, together with 
the dopant nature of adsorbates on those states may be responsible for the large experimental 
variability  that has plagued molecular electronics.\citep{variability1,variability2} 
However, the identification of the key role of this state may enable its detailed control
by deliberate doping. This opens a window of opportunity to progress in this field,
by turning an obstacle into a useful tuning tool. 

 \begin{figure*} 
 \centering    
\includegraphics[width=0.75\textwidth]{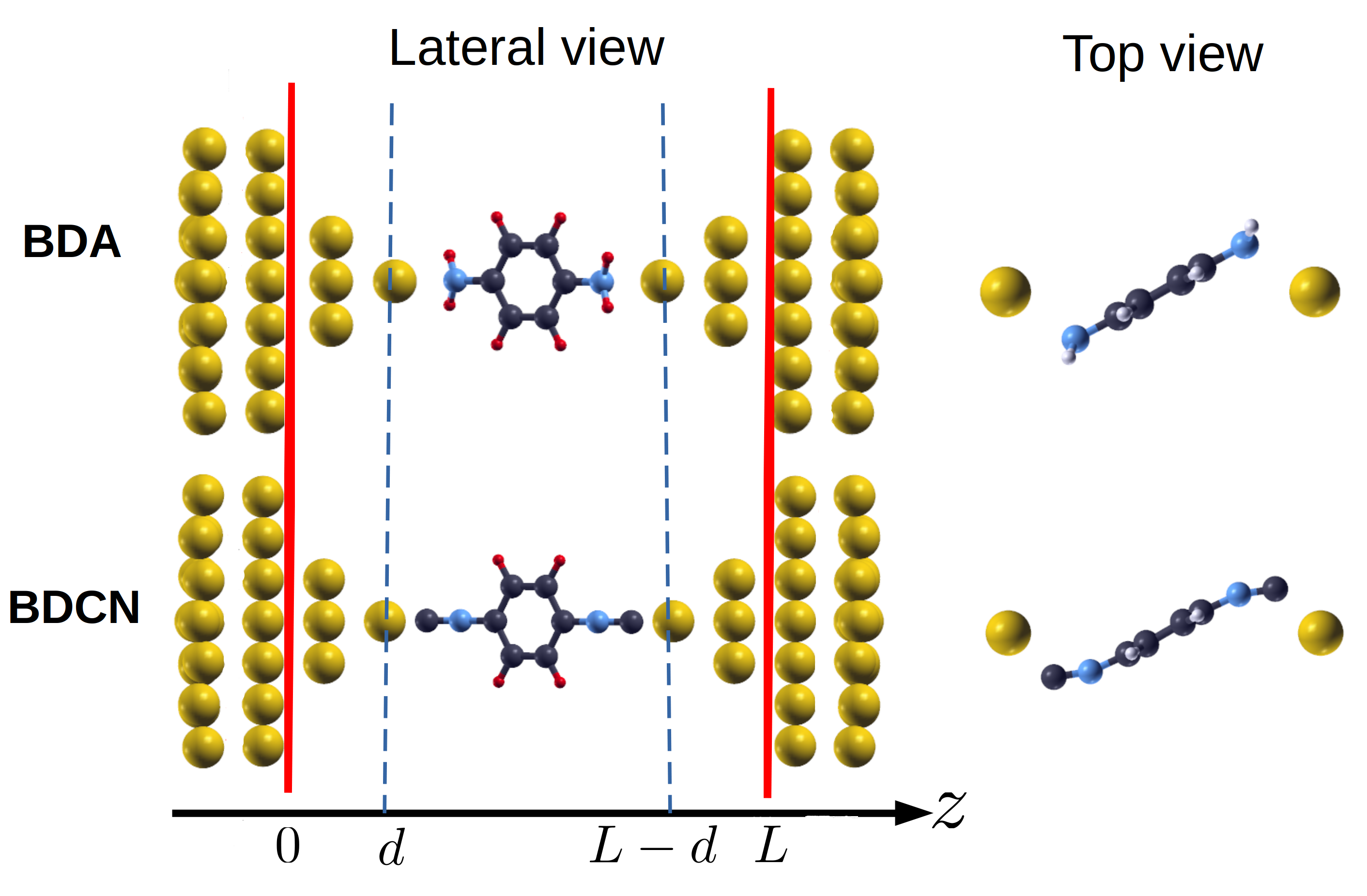}  
\caption{Top and lateral views of the molecular junction structure of a BDA (top) and a BDCN (bottom) molecule bridged between Au leads. Black, blue, red and yellow spheres are used to represent C, N, H and Au atoms, respectively.
The red solid lines indicate the position of the image planes at $z=0$ and $z=L$ for DFT+$\Sigma$, that are located 1 \r{A} from the flat surface towards the molecule. Blue dashed lines represent the position of the tip atom.} 
\label{fig:junction}
\end{figure*} 

A large body of the {\it ab initio} simulations of transport through gold-based junctions performed
in the past has skipped gold 5$d$ states from the valence shell to render the simulations 
computationally manageable. The rationale behind this choice relied on the assumption that
the top of gold 5$d$ bands lie about 1-2 eV below the Fermi energy,
and therefore their impact in the low voltage transport properties of the junction should be negligible. 
However, we find that SSs appear at gold (111) surface simulations only if the 5$d$ states are kept in the valence shell.

We present in this article {\it ab initio} transport simulations of gold (111) junctions to illustrate the relevance of SSs.
We have chosen Benzenediamine (BDA) and 1,4-dicyano-benzene (BDCN) molecules as representative
candidates for the junction barrier, because they have been extensively analysed in the past both experimentally and theoretically.

The article outline is as follows. In the next section, we start with a brief description of the computational methods, with special focus
in reminding the DFT+$\Sigma$ method and the main expressions that we use to compute the conductance $G$ and thermopower $S$
of the simulated junctions. In section 3, we continue with a description of our results and a discussion of how 
they impact in the common lore of the field. A short conclusion closes the article. A thorough description
of the computational details of our simulations is relegated to Appendix A.
\begin{figure*}[t]
  \includegraphics[width=1\linewidth]{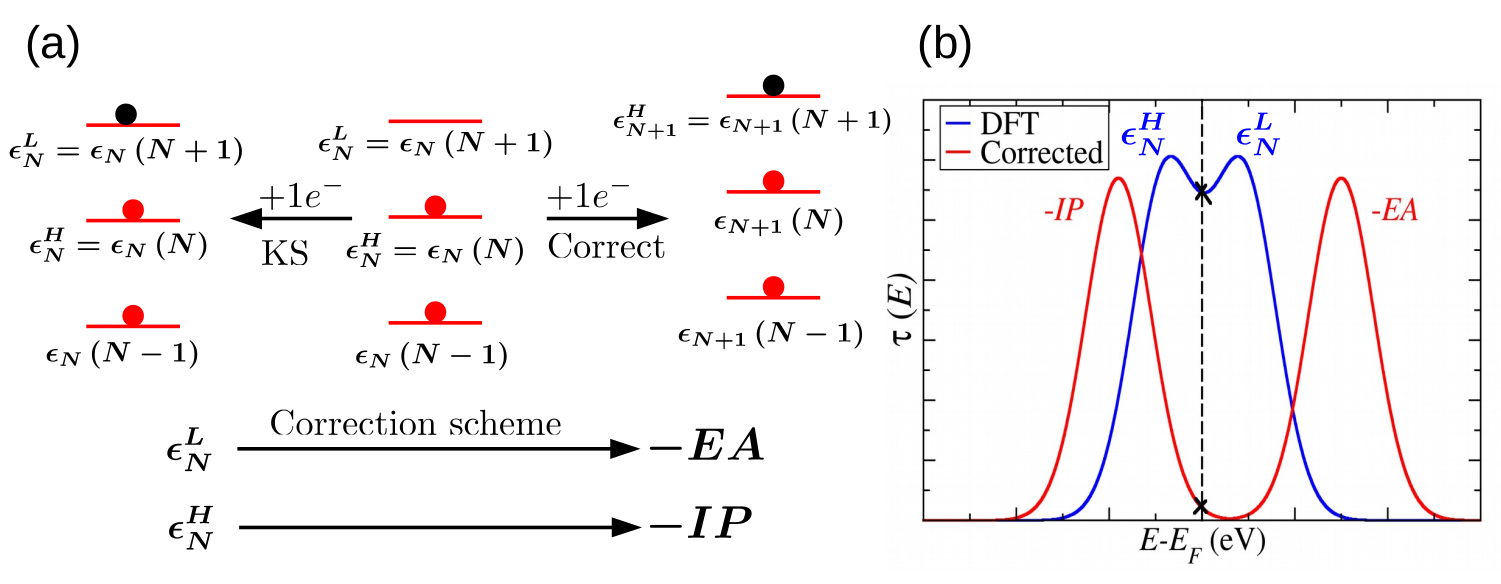}
    \caption{(a) The Kohn-Sham (KS) approach for DFT does not take into account the level shift when an electron is introduced or removed from the system. To correct it we calculate the total energy of the system with one electron added or removed and obtain the ionization potential (IP) and electron affinity (EA). The subscript denotes the total number of electrons in the system while the number between parenthesis represents the energy level (b) Sketch of the transmission function for DFT (blue) and the corrected result (red). Cross symbols give the conductance  value for each method. The DFT+ $\Sigma$ correction shifts the DFT peaks to 
    match the corrected ones.}
\label{fig:scheme}
\end{figure*} 
\section{Computational Details}
We have performed DFT simulations of BDA and BDCN molecules between gold electrodes grown in the (111) direction with the code SIESTA.\citep{siesta} Details of the simulations are relegated to Appendix A.

These simulations are performed in a 2 step process. In the first step, devoted to obtain the relaxed structure of each molecule in the junction, we define a periodic unit cell that includes 2 electrodes of gold formed by 148 atoms each (9 layers of 16 atoms plus a tip of 4 atoms),
bridged by the corresponding molecule.

The second step is aimed to obtaining the electrical and thermoelectrical response of the junctions with the code GOLLUM,\citep{gollum} an implementation of the non-equilibrium Green's function (NEGF) method.\citep{NEGF1,NEGF2} For this purpose, the relaxed structure is connected to another ideal 6 layers of gold on each side to obtain the electronic Hamiltonian $H_{DFT}$ that is latter used by GOLLUM in transport calculations. In Fig. \ref{fig:junction} we show a sketch of the relaxed structure of a BDA and BDCN junction. We consider the left ($L$) and right ($R$) electrodes to be characterized by chemical potentials $\mu_L$ and $\mu_R$ and temperatures $T_L$ and $T_R$. In the limit of small differences $V=(\mu_L-\mu_R)/e$ and $\Delta T=T_L-T_R$, the conductance ($G$) and thermopower ($S$) can be obtained as follows:
\begin{align}
G=G_0\tau(&E_F) \\
S=-\lim_{\Delta T \to 0} &\frac{\Delta V}{\Delta T}\Bigr\rvert_{I=0}=-S_0\frac{\partial \,\mathrm{ln}\big(\tau (E)\big)}{\partial E} \Bigr\rvert_{E = E_F}   \nonumber
\end{align}
where $G_0$ is the conductance quantum, $E_F$ is the Fermi level, $S_0=\pi^2k_B^2T/3e=7.08\,$ eV $\mu$V/K at $T=$300 K and $\tau (E)$ denotes the transmission function for electrons of energy $E$ passing from one electrode to the other. We notice that a large thermopower and thus a larger figure of merit is achieved when the slope of the transmission function is steep near the Fermi energy. We also remark that the sign of the thermopower will be opposite to that of the slope of $\tau(E)$ at the Fermi energy. 

The use of $H_{DFT}$ in the calculation of the transmission function presents two main problems: (1) The Kohn-Sham (KS) approach to DFT does not take into account the level shift induced when an electron is introduced or removed from the system. This implies that the position of the HOMO and LUMO energy levels of the isolated molecule, that govern the transport properties close to the Fermi level, are not calculated correctly (Fig. \ref{fig:scheme} (a)). (2) Screening effects in the electronic states of the molecule due to the presence of a metallic surface might be important,
and they are not included in the description. To solve both problems we introduce the DFT+$\Sigma$ method (Fig. \ref{fig:scheme} (b)). 

To correct the DFT energy levels of the isolated molecule, we introduce the gas phase correction.\cite{sigma2,tipimageplanes} This correction requires the calculation of the total energy of the isolated molecule with one extra or missing electron, $E(N+1)$ and $E(N-1)$, respectively. Then, the total energy of the neutral molecule $E(N)$ is used to compute the ionization potential $IP$ and the electron affinity $EA$ as follows:
\begin{eqnarray}
	IP=E(N-1)-E(N) \\
	EA=E(N)-E(N+1) \nonumber
\end{eqnarray}

Then, the gas phase correction consists in a uniform shift of the KS occupied (and another one for the unoccupied) levels of the isolated, neutral molecule, so that the HOMO (LUMO) state matches the negative IP (EA) (Fig. \ref{fig:scheme} (a)).
\\
\indent Next, we take into account the influence of the metallic electrodes in the electronic states of the molecule when it is inside a junction via the image charge correction. When  a  molecule  approaches  a  metallic surface,  image  charge  interactions  will screen the charge, therefore moving  
the  occupied (unocuppied)  levels  up  (down)  in  energy.\cite{imagecharge1,imagecharge2,imagecharge3} 
To  estimate  the  image  charge  corrections:  (1)  We  obtain  the Hamiltonian $H$ and overlap matrix $S$ 
of the molecule connected to the electrodes from SIESTA. (2) From these matrices we cut  out  the  submatrices 
$H_{mol}$ and $S_{mol}$ spanned  only by  the  basis  functions  on  the  molecular atoms.  (3)  The KS
eigenenergies $\epsilon_{\alpha}$ and eigenvectors $\ket{ \psi_{\alpha} }$ of the molecule in the junction are obtained from 
$H_{mol}\, \ket{\psi_{\alpha}}=\epsilon_{\alpha}\,S_{mol}\,\ket{\psi_{\alpha}}$.  Each eigenstate is written in a basis of $N$ localized orbitals $\ket{\phi_{\mu}}$ as $\ket{\psi_{\alpha}}=\sum_{\mu=1}^{N} c_{\alpha}^{\mu}\ket{\phi_{\mu}}$. 
Each of these localized orbitals can be associated to an atom $a$, so we identify as $a_1, a_2 ... a_{N_a}$ the $N_a$ orbitals centered on atom $a$.
From this decomposition we can define the density matrix of eigenstate $\alpha$ between two atoms $a$ and $b$ as:

\begin{equation}
\rho_{\alpha}^{ab}=
  \begin{pmatrix}
    c_{\alpha}^{a_1}\\
    c_{\alpha}^{a_2}\\
    \vdots\\
    c_{\alpha}^{a_{Na}}
  \end{pmatrix}
  \begin{pmatrix}
    c_{\alpha}^{b_1} & c_{\alpha}^{b_2} &     & \hdots    &     &c_{\alpha}^{b_{Nb}}\\
  \end{pmatrix}
\end{equation}

\noindent 
The charge distribution for a given eigenstate is then:
\begin{equation}
Q_{\alpha} (\vec{r})=\sum_{a}\sum_b \rho_{\alpha}^{ab}\,\,\, S_{mol}^{ba}\,\,\, \delta (r-\vec{R_a})\\
\end{equation}
\noindent 
The image charge energy for a point charge distribution $Q_{\alpha}(\vec{r})$ placed between two
image planes located at $z=0$ and $z=L$ is:
\begin{align}
\begin{split}
 	&\Delta_{\alpha}=\frac{1}{8\pi\epsilon_0}\sum_{a,b} Q_{\alpha} (r_{a})Q_{\alpha} (r_{b})\Bigg( \Bigg.\sum_{n=1}^{\infty} 		\frac{1}{\sqrt{(z_a+z_b-2nL)^2+r_{ab}^2}}+\\&+\frac{1}{\sqrt{(z_a+z_b+2\,(n-1)L)^2+r_{ab}^2}}-\\&-\frac{1}{\sqrt{(z_a-z_b+2nL)^2+r_{ab}^2}}-\frac{1}{\sqrt{(z_a-z_b-2nL)^2+r_{ab}^2}} \Bigg.\Bigg)
\end{split}
\end{align}
\noindent
where $x_a, y_a, z_a$ are the cartesian coordinates of atom $a$, with $z$ being the transport direction, and $r_{ab}$ =$\sqrt{(y_{a}-y_{b})^2+(x_{a}-x_{b})^2}$. 

Screening effects due to a flat metallic electrode can be described as a classical, semi-infinite conductor with a flat surface located at a distance $\sim 1$ \r{A} outside the last 
atomic layer.\cite{flatimageplanes,flatimageplanes2} However, a tip terminated electrode has only one atom in its last layer, and therefore its equivalence to a flat surface is questionable. To solve this problem, we will place the image plane $\sim 1$ \r{A} towards the molecule from the flat surface (last layer not taking into account the pyramid).

Small changes in the position of this plane lead to negligible changes in the correction.
In Fig. \ref{fig:junction} the difference between both approaches
is shown. Similarly to the application of the gas phase correction, we shift all the occupied (unoccupied) levels by the correction $\Delta_o$ ($\Delta_u$) corresponding to the charge distribution of the HOMO (LUMO) state. Including both corrections, we obtain the corrected energy spectrum of the occupied ($\epsilon_{oc}^c$) and unoccupied ($\epsilon_{un}^c$) states as:
\begin{eqnarray}
	 \epsilon_{oc}^c=\epsilon_{oc}-\epsilon_{HOMO}-IP+\Delta_o\\
	 \epsilon_{un}^c=\epsilon_{un}-\epsilon_{LUMO}-EA-\Delta_u \nonumber
\end{eqnarray}

We then obtain a corrected molecular Hamiltonian $H_{mol}^c$ as:
\begin{equation}
H_{mol}^c=\sum_{\alpha\in oc} \epsilon_{\alpha}^c\ket{\psi_{\alpha}}\bra{\psi_{\alpha}}+\sum_{\beta\in un} \epsilon_{\beta}^c\ket{\psi_{\beta}}\bra{\psi_{\beta}}
\end{equation}

This corrected term substitutes $H_{mol}$ inside $H_{DFT}$ to obtain the corrected $H_{DFT+\Sigma}$ Hamiltonian used to determine the transport properties of the junctions.

\section{Results} 

\begin{figure}[t]
  \includegraphics[width=1\linewidth]{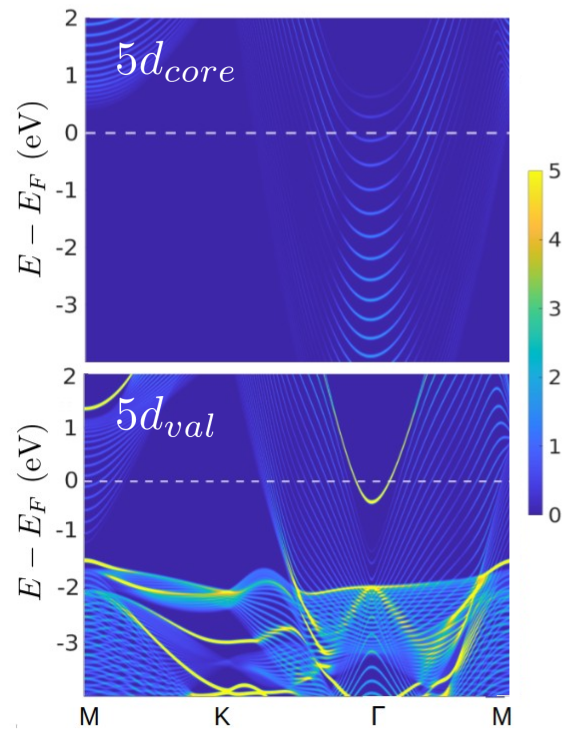}
    \caption{Band structure of clean gold (111) surface projected on the surface atom along the k-path MK$\Gamma$M for the system where $5d$ orbitals are included explicitly in the core (top panel) or in the valence (bottom panel) shell. The white dashed line denotes the position of the Fermi energy. The Shockley state can be seen in the $5d_{val}$ case at $\sim$0.5 eV below the Fermi energy around the $\Gamma$ point.}
\label{fig:PDOSK}
\end{figure}

\begin{figure*}[t]
  \includegraphics[width=1\linewidth]{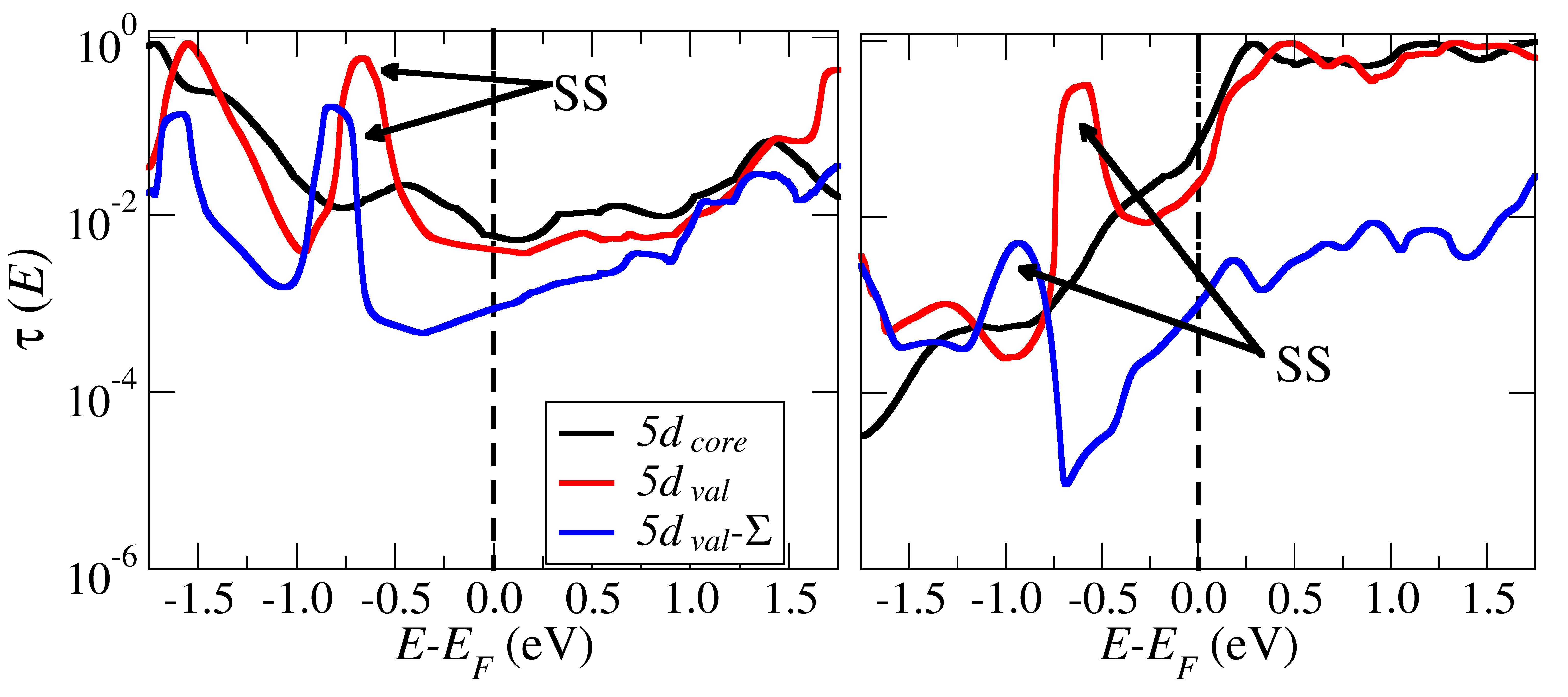}
    \caption{Transmission function of BDA (left panel) and BDCN (right panel). Calculations with the $5d$ orbitals included in valence (core) shell are denoted with $5d_{val}$ ($5d_{core}$). Blue line shows the transmission function with the $\Sigma$ correction. The SS can be seen when $5d$ orbitals are in the valence while it disappears for the $5d$ orbitals in the core shell. The intersection between the graphs and the vertical black line gives the low-voltage conductance $G=G_0 \tau (E_F)$}
\label{fig:transmisiones}
\end{figure*}

In this section we show the relevance of the gold SS in the calculated electric and thermoelectric properties of molecular junctions, illustrated by the examples of BDA and BDCN. We start our discussion with the analysis of the energy position and appearance of the SS at the gold (111) surface. For this purpose, we simulate a gold slab grown in the (111) direction, with its experimental lattice constant of $a=4.07$ \r{A}. We perform the simulations including the 5$d$ orbitals of Au in the valence (labeled as $5d_{val}$) or keeping them in the core (labeled $5d_{core}$). For both cases, we compute the band structure of the gold slab, and project it over the surface atoms (Fig. \ref{fig:PDOSK}). The SS, manifested by  a parabola around the $\Gamma$ point with a large contribution from the surface orbitals,
appears only for the $5d_{val}$ case. Moreover, its correct position at $\sim$ 0.5 eV below the Fermi level\cite{Papior} is only reproduced if we consider the experimental lattice constant (see Appendix B for more details). Our assumption is that the presence of this SS, being so close to the Fermi level, will play an important role in the electric and thermoelectric properties of gold-based junctions.
To confirm this hypothesis, we compute the transmission function $\tau(E)$ for the BDA and BDCN junctions using the $5d_{core}$ and $5d_{val}$ DFT Hamiltonians. We show our results in Fig. \ref{fig:transmisiones}, where we also include the results with the $\Sigma$ correction for the $5d_{val}$ case, labeled $5d_{val}$-$\Sigma$.

Unlike what many other transport codes do,\citep{transiesta,kwant} GOLLUM does not include a fictitious imaginary part in the Green's function calculation, but obtains it exactly. This leads to transmission curves with more structure than other methods, since the imaginary part acts as a softing parameter of the transmission.
We find that the results are clearly dependent on the Hamiltonian used. First of all, the presence of the SS in the gold surface of the $5d_{val}$ cases lead to a well defined peak in the transmission below the Fermi level. For the $5d_{val}$ calculations, we find this peak at an energy close to 0.5 eV below the Fermi level, but slightly different for each of the two molecules. This molecule-dependent shift of the SS respect to its position in a clean gold surface is due to its hybridization with the states of the molecule. This hybridization also implies that the correct position of the SS is only obtained after the application of the $\Sigma$ correction, even though this correction has no direct effect over the Hamiltonian of the electrodes. This suggests that changing the anchoring group of the molecule could allow to tune the position of the SS.

We discuss next the impact of the SS in the transmission function close to the Fermi level. To complement Fig. \ref{fig:transmisiones}, Table \ref{table:results} summarizes the electric and thermoelectric properties of the junctions based on $\tau(E_F)$. The transmission through BDA is HOMO-dominated and has negative slope at the Fermi energy (and thus positive thermopower)  for the $5d_{core}$ and $5d_{val}$ cases. On the other hand, for the $5d_{val}$-$\Sigma$ case, the Shockley state moves to lower energies ($\sim$ 0.2 eV), thermopower becomes negative and conductance drops by a factor of 5 respect to the $5d_{val}$ case.
For BDCN, the transmission is LUMO-dominated and has positive slope for all cases. Including the $\Sigma$ correction shifts the position of the SS state almost 0.5 eV, and the conductance is then reduced by more than an order of magnitude. However, the similar slope of the transmission in logarithmic scale leads to similar
values of the thermopower for all cases. 
\begin{figure*}[t]
  \includegraphics[width=0.9\linewidth]{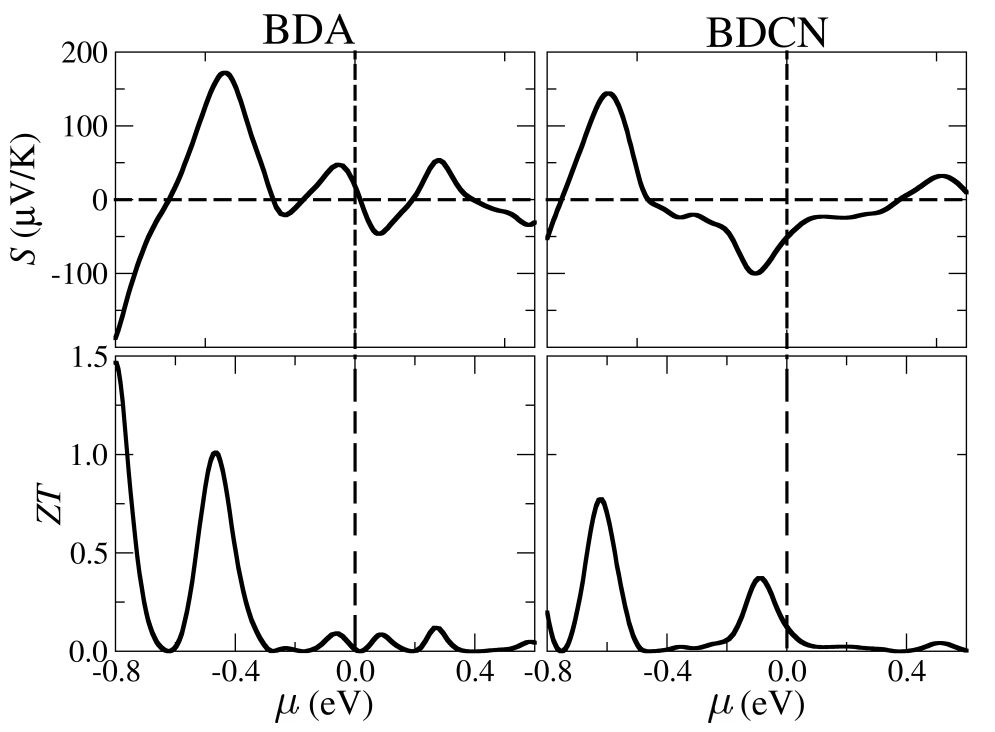}
    \caption{Thermopower and figure of merit for different chemical potentials $\mu$ for BDA (left column) and BDCN (right column). All calculations were done with $5d$ orbitals in valence. The vertical dashed line denotes zzero-gating in which all the previous calculations were done. The Shockley state can be clearly seen at $\sim$ 0.5 eV below the Fermi energy. }
\label{fig:SandZT}
\end{figure*} 

Lastly, we remark that some of the features in the transmission function around -1.5 eV and 1.5 eV reflect the local density of states at the tip Au atom and are not related to the molecular levels, with only small shifts due to the hybridization between the states of the molecule and those of the tip. Since the gold atoms are always treated at DFT level in our calculations, these features appear at approximately the same energies in the DFT and DFT+$\Sigma$ spectra. 

Beyond these results, the well defined peak associated to the SS covers a huge range of values of $\tau (E)$, but also of its derivative. Then, if we could tune its position, eventually making it pass through the Fermi level, we could tune electric and thermoelectric response at will. As it has already been discussed, changes in the anchoring group of the molecule is a potential way to tune the position of the SS. We consider other tuning options. As we show in detail in the Appendix B, changes in the gold lattice constant have a huge impact in the position of the SS. However, to change the lattice constant of gold enough so that the shift in the SS position is noticeable, the pressure required is in the order of several GPa, and therefore its experimental realization results challenging. Another option would be to shift the chemical potential of gold, via a gate voltage, chemical doping of the surface, or other methods, which are more plausible experimentally.
\begin{table}
\centering
\setlength{\extrarowheight}{7pt}
\resizebox{\linewidth}{!}{%
\begin{tabular}{c|ccc|ccc}
    \multirow{2}{*}{Molecule} &
      \multicolumn{3}{c|}{G ($10^{-3}G_0$)} &
      \multicolumn{3}{c}{S ($\mu$V/K)}  \\
    & $5d_{core}$ & $5d_{val}$ & $5d_{val}$-$\Sigma$& $5d_{core}$ & $5d_{val}$ & $5d_{val}$-$\Sigma$ \\
    \hline
    BDA & 7.8 & 3.9 & 0.8 & 2.7 & 3.3 & -7.5 \\
    \hline
    BDCN & 62 & 22 & 0.1 & -73.8 & -46.2 & -46.9 \\
  \end{tabular}}
  \caption{Calculated conductance and thermopower for BDA and BDCN junctions. The corresponding transmission functions are plotted in Fig. \ref{fig:transmisiones}. }
\label{table:results}
\end{table}
In Fig. \ref{fig:SandZT} the thermopower and the figure of merit as a function of a shift in the chemical potential are plotted. Starting by BDA, two very interesting feature can be seen in the graphs (1) small changes near zero-gating (where all the previous calculations were done) change the behavior of the system completely: we can go from positive to negative thermopower by moving the chemical potential less than 0.1 eV. Apart from that, a change in ZT from almost 0 to $\sim$0.1 can also be seen for that shift. The sensitivity of the thermoelectric properties to a change in chemical potential in the simulation may indicate that this system is experimentally sensitive to changes (gate voltage, doping...), wether intentional or unavoidable. Apart from that, we can exploit this sensitivity to build a molecular switch (two different states separated by a barrier that we can control) for different applications. (2) If a shift of 0.5 eV is experimentally possible, we would move the Shockley state to just below the Fermi energy enhancing the thermoelectrical properties of the junction from a thermopower of 30 $\mu$V/K and a ZT of 0 to 300 $\mu$V/K and 0.9. This result will have a large impact for the molecular electronics community as many different strategies have been carried out in the later years to achieve a Seebeck coefficient in the order of 500 $\mu$V/K.

Turning now to BDCN Fig. \ref{fig:SandZT}, we can see that small changes in the chemical potential near the equilibrium point change the value of the Seebeck coefficient by a factor of two but not its sign (the change in behaviour would not be as abrupt as in BDA). On the other hand, with a smaller shift (in the order of 0.1 eV) we can see a significant change in the ZT going from 0.1 to almost 0.4: this junction will be more sensitive to doping and thus it may be a better candidate for a molecular switch. As in the previous case, the Shockley state can be clearly seen $\sim$0.5 eV below the Fermi energy with a smaller ZT value that for BDA.
\begin{figure*}[t]
\centering    
\includegraphics[width=1\textwidth]{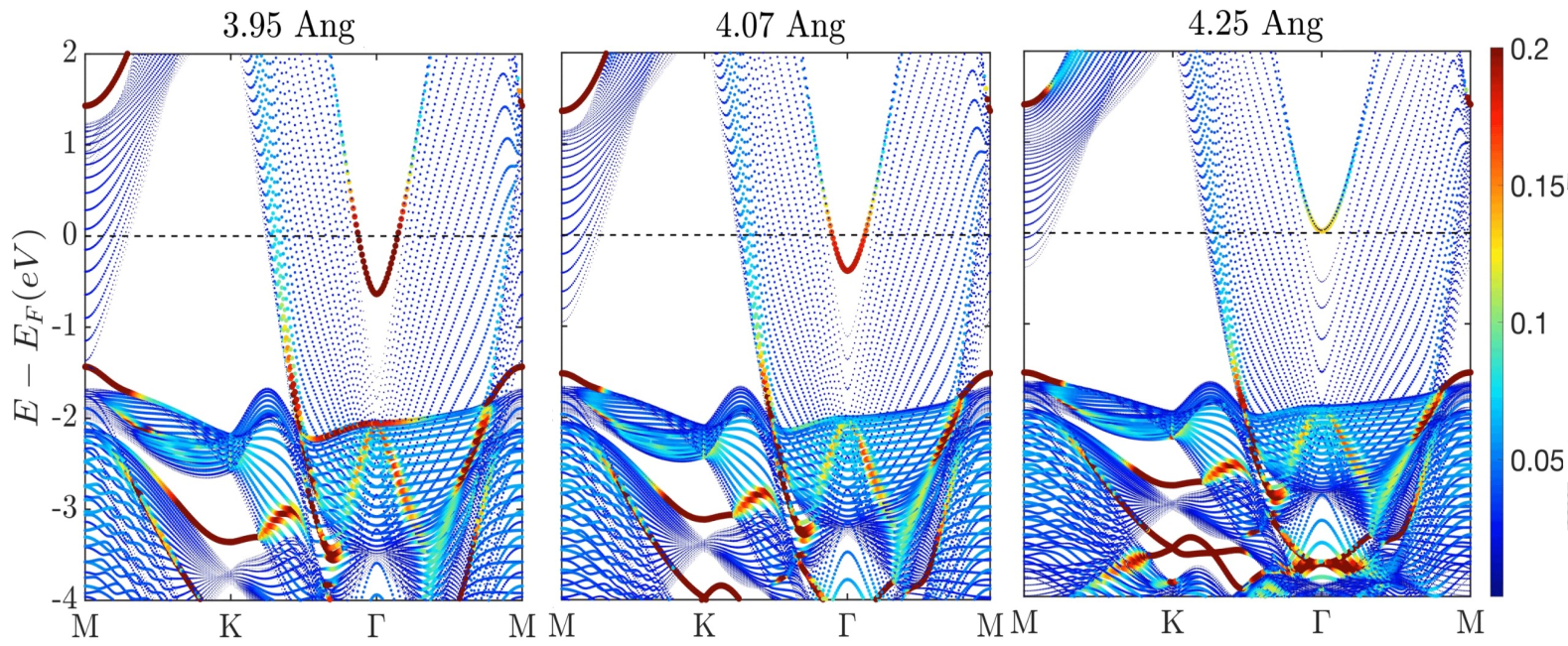}  
\caption{Band structure along the MK$\Gamma$M for different lattice constants. For the experimental lattice constant $a=4.07$ \r{A}, in the center pannel, we see the SS 0.5 eV below the Fermi energy as shown in the transport calculations. } 
\label{fig:bandas_pressure}
\end{figure*} 
\section{Conclusions}
In summary, using the DFT+$\Sigma$ correction scheme, we have
demonstrated how Shockley surface states can be properly
incorporated in {\it ab initio} electron transport calculations going
beyond DFT. This result may motivate further experiments in the molecular electronics community as most electrodes are made of gold
(111). We show how SSs can change the conductance and
the thermopower significantly: it leads to larger experimental values of the figure of merit and will be instrumental to find new and more efficient thermoelectric devices. Lastly, we study how different physical parameters such as pressure and chemical doping may shift the position of the Shockley state and how this can be exploited to fine tune the properties of the junction.

\section*{Author Contributions}

S.Sánchez-González performed the simulations with assistance of A. García-Fuente and J. Ferrer. A. García-Fuente and J. Ferrer supervised the research and provided essential contributions to interpreting the results. All co-authors assisted in writing the manuscript. 

\section*{Conflicts of interest}
There are no conflicts to declare.

\section*{Acknowledgements}
This work was supported by Ministerio de Ciencia, Innovacion y Universidades, Agencia Estatal de Investigacion, Fondo Europeo de Desarrollo Regional via the Grant PGC2018-094783, and by 
Centro de Investigacion en Nanomateriales y Nanotecnologia (CINN) with a JAE Intro ICUs Grant.
S.Sánchez-González acknowledges the grant PRE2019-088087 from the Spanish Ministry of Science and Universities for funding.
\begin{figure}[h!]
\centering    
\includegraphics[width=1\linewidth]{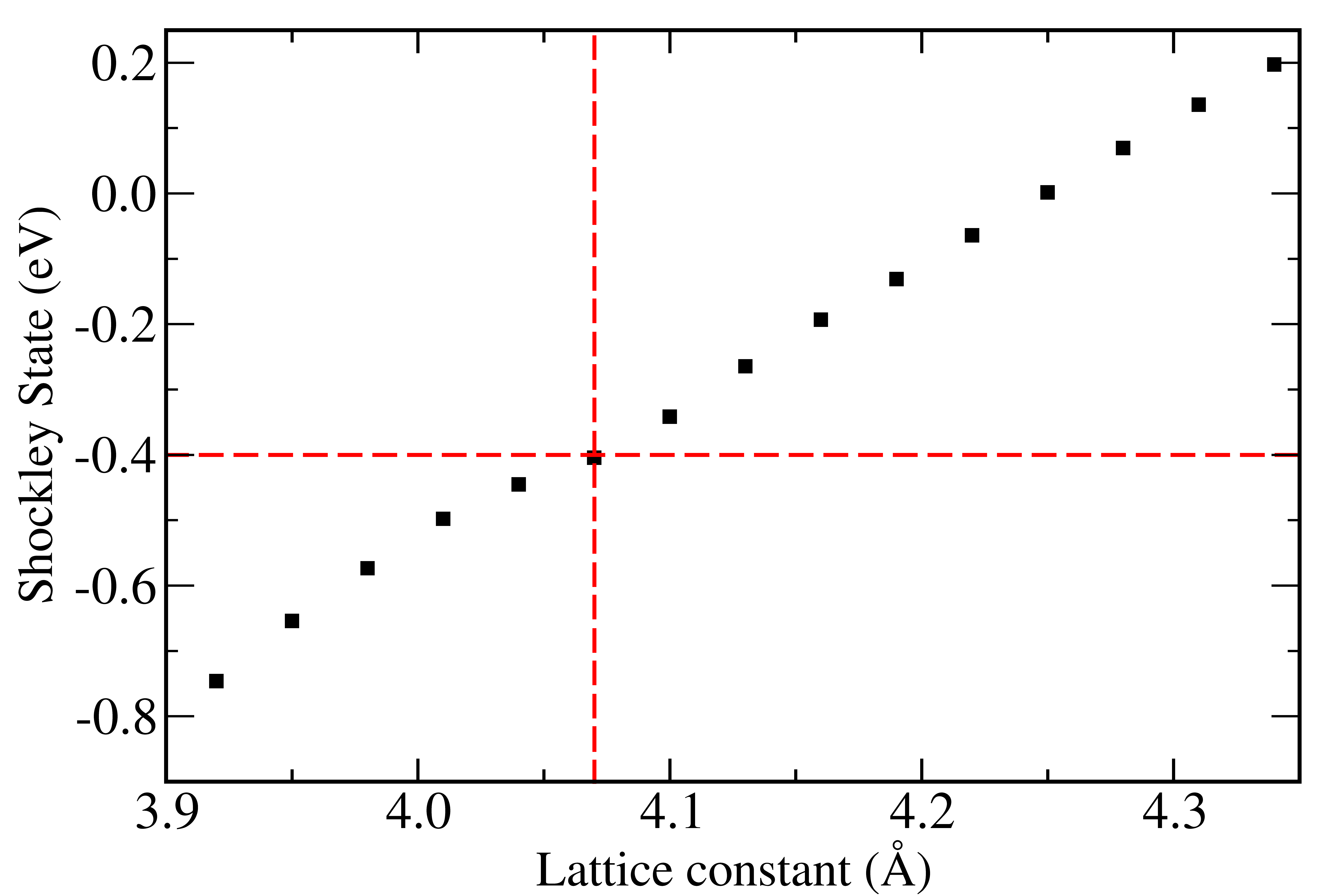}  
\caption{SS position dependance on the lattice constant. The vertical dashed line represent the experimental lattice constant $a=4.07$ \r{A}} 
\label{fig:SS_position}
\end{figure}

\section*{Appendix A: Computational Details}
DFT calculations were performed using the code SIESTA \citep{siesta}, with the PBE parametrization of the exchange-correlation potential.\citep{PBE} SIESTA expands the wavefunctions of the valence electrons in a basis of pseudo-atomic localized orbitals, that extend up to a cutoff radius $r_c$. We use a DZP basis in all our calculations, with the following orbitals included in the valence, and their corresponding values of $r_c$ in Bohr units for each atom: Gold: 8.009, 6.008 (6s), 9.075 (6p) and 4.078, 2.946 (5d). Hydrogen: 7.026, 4.104 (1s). Carbon: 7.086, 3.519 (2s), 7.086,3.793 (2p). Nitrogen: 7.057, 2.942 (2s) and 7.057, 3.171 (2p). Calculations done with the Au 5d states kept in the core use the same basis orbitals for the 6s and 6p states. Interaction of the valence states with the core electrons is included via pseudopotentials obtained from the work by Rivero et al. \citep{pseudos} Real space integrals were computed in a mesh defined by a cutoff of 400Ry

\section*{Appendix B: Effect of pressure on the SS position}
The effect of changing the lattice constant of bulk Au (equivalent to applying pressure experimentally) in the position of the SS has been studied. In Fig. \ref{fig:bandas_pressure} we show the band structure along the MK$\Gamma$M path for different lattice constants. The SS can be seen as the bottom of a parabola at the $\Gamma$ point. As the pressure increases (from right to left pannels) the SS is shifted to lower energies going from an unoccupied state to an occupied one. These results can be better seen in Fig. \ref{fig:SS_position}. Unfortunately, the pressures needed to shift the SS an appreciable energy are in the order of several GPa, making this tuning tool impossible experimentally as the electrodes will break under so much stress.

\bibliography{paper.bib}

\end{document}